\documentclass[prl,showpacs,10pt,preprint,floatfix,groupedaddress]{revtex4-1}
\usepackage{graphicx}
\usepackage{amssymb}
\usepackage{amsmath}

\begin{document}
\preprint{YITP-SB-11-32}

\title{Heavy quarkonium production and polarization} 

\author{Zhong-Bo Kang$^{1}$, Jian-Wei Qiu$^{2,3}$ and George Sterman$^3$}

\affiliation{$^1$RIKEN BNL Research Center, 
                 Brookhaven National Laboratory, 
                 Upton, NY 11973, USA}
\affiliation{$^2$Physics Department,
                 Brookhaven National Laboratory,
                 Upton, NY 11973, USA}
\affiliation{$^3$C.N. Yang Institute for Theoretical Physics,
                 Stony Brook University,
                 Stony Brook, NY 11794, USA}

\date{\today}

\begin{abstract}
We present a perturbative QCD factorization formalism for the production of heavy quarkonia of large transverse momentum $p_T$ at collider energies, which includes both the leading power (LP) and next-to-leading power (NLP) contributions to the cross section in the $m_Q^2/p_T^2$ expansion for heavy quark mass $m_Q$.  We estimate fragmentation functions in the non-relativistic QCD formalism, and reproduce the bulk of the large enhancement found in explicit NLO calculations in the color singlet model.   Heavy quarkonia  produced from NLP channels prefer longitudinal polarization.   
\end{abstract}

\maketitle

%

\noindent
{\it{ Introduction}}\,---\,More than thirty-five years after the discovery of the $J/\psi$~\cite{jpsi-disc}, the production of heavy quarkonia remains a key subject in strong interaction physics \cite{Brambilla:2010cs}.  The inclusive production of pairs of charm or bottom quarks, with masses $m_Q\gg \Lambda_{\rm QCD}$, is an essentially perturbative process, while the subsequent evolution of the pair into a physical quarkonium is  nonperturbative.  Different treatments of the transformation from  heavy quark pair to bound quarkonium are given in various formalisms, most notably, the color singlet model (CSM), the color evaporation model (CEM), and non-relativistic QCD (NRQCD).  The current status of theory and experiment has been summarized very recently in Ref.~\cite{Brambilla:2010cs}.

For the NRQCD formalism \cite{Caswell:1985ui,Bodwin:1994jh},  small, color-octet production matrix elements can provide good fits to high $p_T$ inclusive hadron collider cross sections for $J/\psi$ and $\Upsilon$, but a complete description remains elusive.  Polarization in particular remains a challenge, along with the surprisingly high rate of associated production in electron-positron annihilation \cite{Brambilla:2010cs}.  Although plausible arguments for the use of NRQCD at leading power ($p_T^{-4}$) have been around since the beginning of the formalism \cite{Bodwin:1994jh}, issues of gauge invariance and infrared cancellation are still not completely settled \cite{Nayak:2005rw,Nayak:2007mb}.  Meanwhile, large next-to-leading order (NLO) and potentially large next-to-next-to leading order (NNLO) corrections to high-$p_T$ cross sections \cite{Campbell:2007ws} in the CSM have attracted attention.  The size of these color singlet cross sections seems to upset long-held expectations for gluon fragmentation/color octet dominance, and indeed, they appear in diagrams that fall off like $p_T^{-6}$ in $d\sigma/dp_T^2$, compared to the leading, $p_T^{-4}$ behavior associated with gluon fragmentation.  These developments suggest that we must widen the formalism for quarkonium production beyond leading power.  

\begin{figure}[!htp]
	\centering
	\includegraphics[width=0.4\columnwidth]{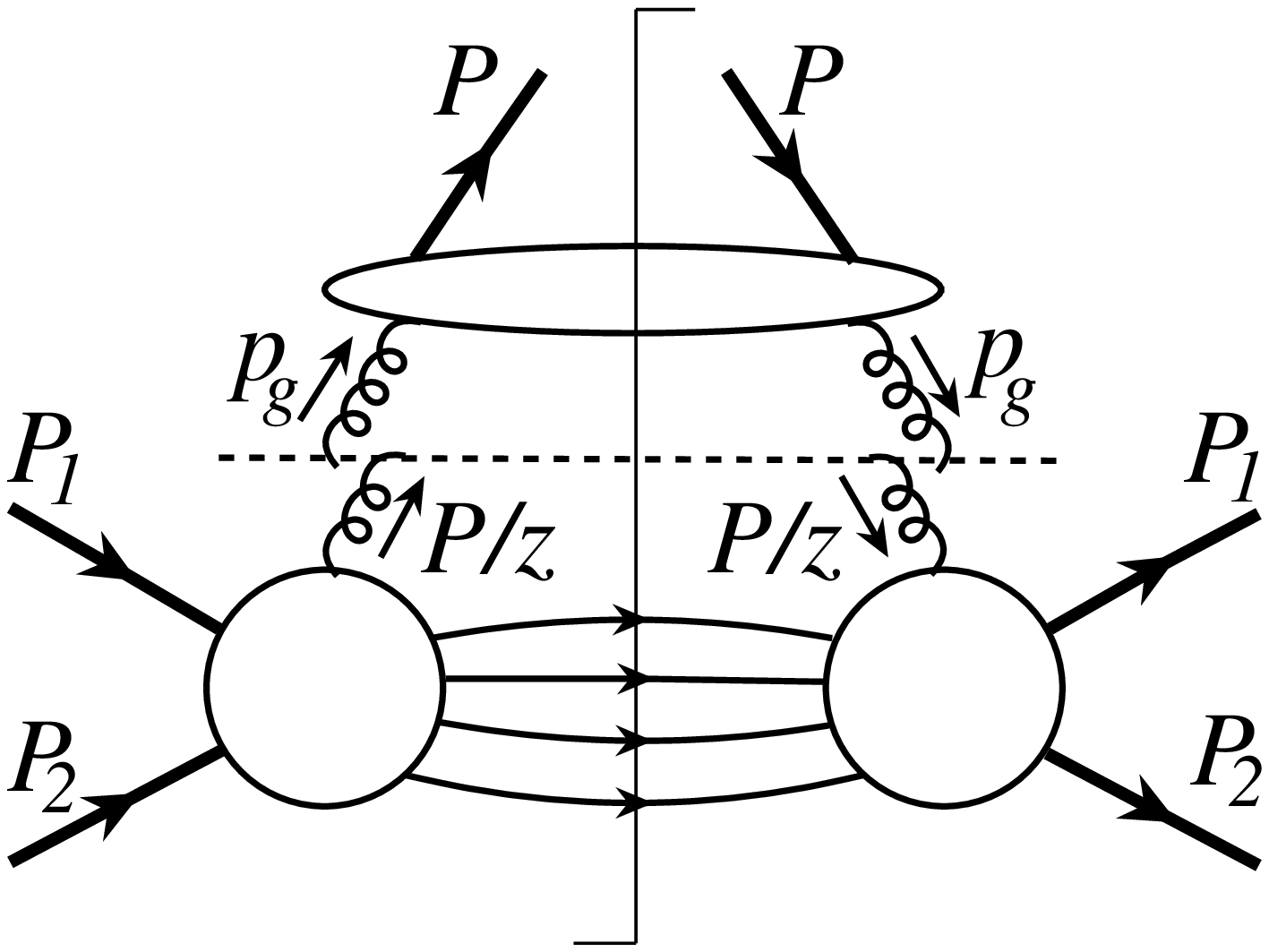}
	\hskip 0.1\columnwidth
    \includegraphics[width=0.4\columnwidth]{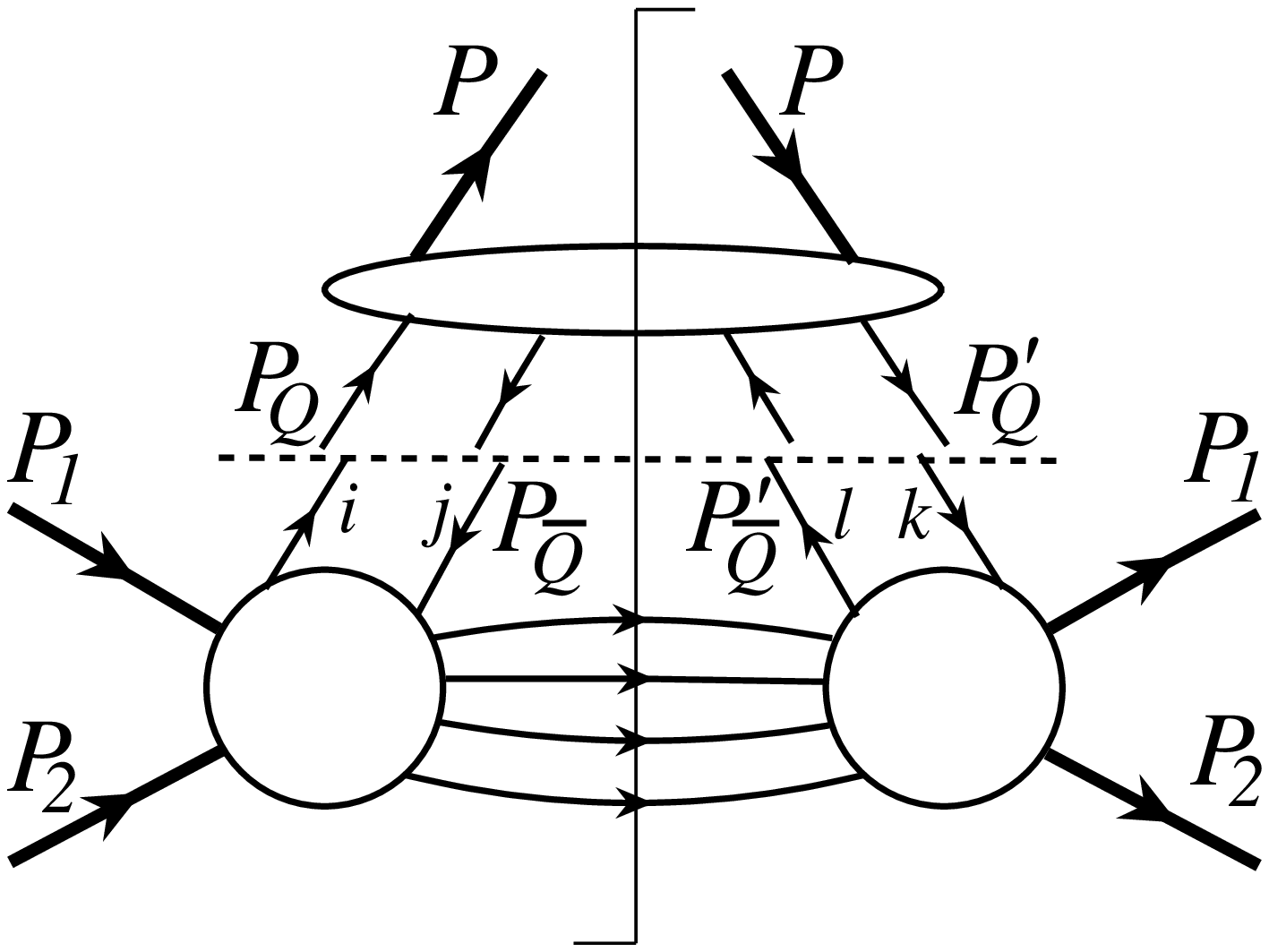}
	\caption{$J/\psi$ production through a single parton (a gluon) fragmentation (left) and fragmentation of a heavy quark pair (right).}
	\label{fig:Jpsi}
\end{figure}

In this letter, we emphasize the expansion of the production rate of the heavy quark pairs first in the large scale $p_T$ and then in the coupling constant $\alpha_s$.   We present a perturbative QCD factorization formalism, accurate to the first nonleading power in $m_Q^2/p_T^2$ that incorporates both leading power gluon fragmentation and direct production of heavy quarks at short distances with subsequent fragmentation, as illustrated in Fig.\ \ref{fig:Jpsi} \cite{Kang:2011zz}. These figures are shown in cut diagram notation, in which the amplitude and complex conjugate are combined into a forward scattering diagram in which the final state is identified by a vertical line.  When $p_T\gg m_Q$, the rate to produce a gluon (or in general a single parton) at distance scale $1/p_T$, which fragments into a heavy quark pair and thence into a physical quarkonium, is characterized by a $p_T^{-4}$ behavior.  We refer to this as the leading power (LP) contribution.  The perturbative production of a collinear heavy quark pair directly at the short distance scale $1/p_T$ is suppressed by $1/p_T^2$ relative to the production rate of a single parton at the same $p_T$.   We therefore refer to it as a next-to-leading power (NLP) contribution.  As we will see below, however, the probability for a such heavy quark pair to evolve into a heavy quarkonium is naturally enhanced compared to that of a single parton.  This can promote the NLP channel to phenomenological interest, despite its suppression by two powers of $p_T$.  

The physical heavy quarkonium is likely formed long after the heavy quark pair was produced \cite{Brodsky:1988xz}.  Both the LP and NLP production channels in Fig.~\ref{fig:Jpsi} can therefore be considered as fragmentation contributions \cite{Brambilla:2010cs}.  Similar to factorization at LP \cite{Nayak:2005rw}, the NLP contribution to the production rate is factorized into perturbatively calculable short-distance coefficient functions for producing the heavy quark pair convoluted with non-perturbative but universal long-distance fragmentation functions for the pair to become a heavy quarkonium \cite{Kang:2011zz}.  The universality of the fragmentation functions can in principle be tested  when we compare the data of heavy quarkonium production from processes with different short-distance coefficients.  As a consequence of this perturbative factorization, the short-distance coefficient functions at both the LP and NLP capture the dynamics at the distance scale $1/p_T$, and are insensitive to the details of the produced heavy quarkonium states.  
 
%

\noindent
{\it{ Factorization formalism}}\,---\,For the production of a heavy quarkonium state $H$ of momentum $p$, $A(P_1)+B(P_2)\to H(p) +X$, the leading contribution from the channels in Fig.~\ref{fig:Jpsi} can be summarized in an extended factorization formula \cite{Kang:2011zz,KQSfac:2011}, given schematically by 
\begin{eqnarray}
\label{eq:pqcd_fac}
d\sigma_{A+B\to H+X}(p)
&\approx &
\\
&\ & \hspace{-20mm}
\sum_{f} 
d\hat{\sigma}_{A+B\to f+X}(p_f=p/z)
\otimes D_{H/f}(z,m_Q)
\nonumber\\
&\ & \hspace{-35mm}
+
\sum_{[Q\bar{Q}(\kappa)]}
d\hat{\sigma}_{A+B\to [Q\bar{Q}(\kappa)]+X}(p(1\pm\zeta)/2z,p(1\pm\zeta')/2z)
\nonumber\\
&\ & \hspace{-10mm}
\otimes {\cal D}_{H/[Q\bar{Q}(\kappa)]}(z,\zeta,\zeta',m_Q) \, ,
\nonumber
\end{eqnarray}
where the factorization scale dependence is suppressed. The first (second) term on the right-hand side gives the contribution of LP (NLP) in $m_Q/p_T$. In the first term, $d\hat\sigma_{A+B\to f+X}(p_f)$ is the cross section to produce an on-shell parton of flavor $f$ at short distances, which contains all of the information about the incoming state and includes convolutions with parton distributions when $A$ or $B$ is a hadron.   The sum over $f$ runs over all parton flavors, and $D_{H/f}(z,m_Q)$ is the fragmentation function for a heavy quarkonium state $H$ from parton $f$ with momentum fraction $z$ \cite{Nayak:2005rw}.  For $H$ a $J/\psi$ or other heavy quarkonium, the dominant channel at hadron colliders is gluon fragmentation, $f=g$.   The second term on the right-hand side in Eq.~(\ref{eq:pqcd_fac}) is suppressed by $p_T^{-2}$ relative to the first, and the quark-pair fragmentation function, ${\cal D}_{H/[Q\bar{Q}(\kappa)]}(z,\zeta,\zeta',m_Q)$, has units of mass squared, which are compensated by large invariants from the hard-scattering function, $d\hat{\sigma}_{A+B\to [Q\bar{Q}(\kappa)]+X}(p(1\pm\zeta)/2z,p(1\pm\zeta')/2z)$, which describes production of an on-shell, collinear heavy quark pair.  The momentum fractions $z$, $\zeta$ and $\zeta'$ are defined as
\begin{eqnarray}
p_Q^+ &=& p^+\, \frac{1+\zeta}{2z}\, , \quad\quad 
p_{\bar Q}^+ = p^+\, \frac{1-\zeta}{2z}\, ,
\nonumber \\
p_Q^{'+} &=& p^+\, \frac{1+\zeta'}{2z}\, , \quad\quad 
p_{\bar Q}^{'+} = p^+\, \frac{1- \zeta'}{2z}\, .
\label{eq:zzetadef}
\end{eqnarray}
By analogy to the single-parton case, $z$ measures the fractional momentum of the pair carried by the observed quarkonium in this leading region, which is the same on both sides of the cut in Fig.\ \ref{fig:Jpsi}. Parameters $\zeta$ and $\zeta'$ characterize the sharing of the pair's momentum between the heavy quark and antiquark on either side of the cut in the figure.   In principle, these need not be the same.  The $\otimes$ in Eq.~(\ref{eq:pqcd_fac}) represents the convolution over the partons' momentum fractions \cite{KQSfac:2011}. 

In Eq.~(\ref{eq:pqcd_fac}), the perturbative hard parts at both the LP and NLP capture the QCD dynamics at distance scale of $1/p_T$, are independent of heavy quark mass, and are the same for the production of all heavy quarkonium states.   For the LP, the hard parts are effectively the same as those for inclusive production of any single hadron at high $p_T$ and are available to both leading order (LO) and next-to-leading order (NLO) in $\alpha_s$ \cite{Aversa:1988vb}.  

\begin{figure}[!htp]
\noindent
\begin{minipage}[c]{0.32\columnwidth}
\noindent
        \includegraphics[width=0.85\columnwidth]{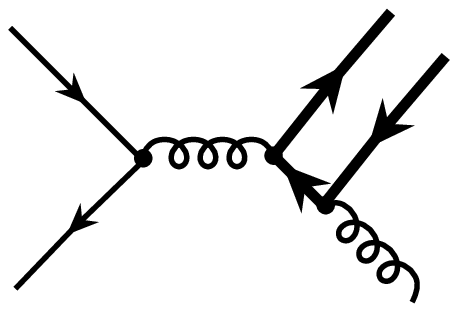}\\
        (a)
\end{minipage}
\begin{minipage}[c]{0.32\columnwidth}
\noindent
        \includegraphics[width=0.85\columnwidth]{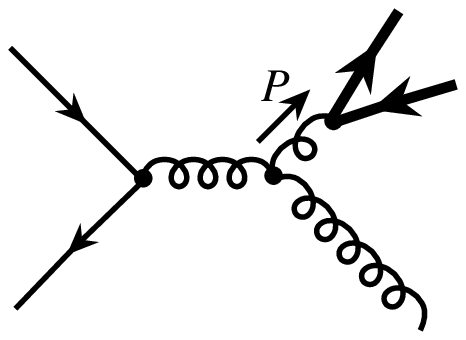}\\
        (b)
\end{minipage}
\begin{minipage}[c]{0.32\columnwidth}
\noindent
        \includegraphics[width=0.85\columnwidth]{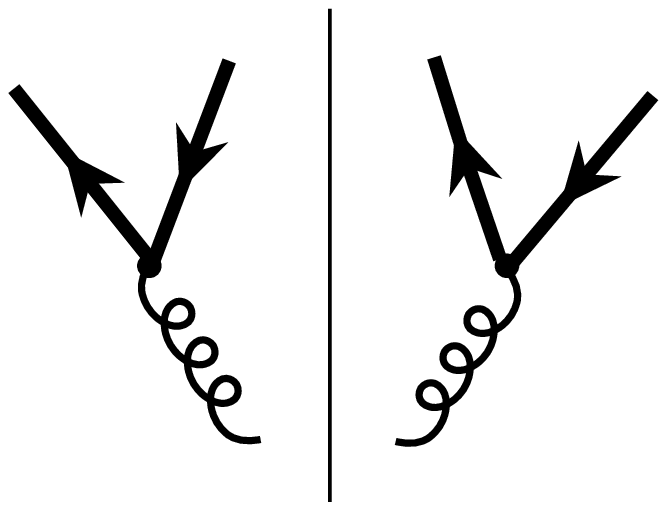}\\
        (c)
\end{minipage}
\caption{Sample Feynman diagrams for $q+\bar{q}\to [Q\bar{Q}(\kappa)](p)+g$ subprocess.  The diagram in (a) contributes to the short-distance coefficient at NLP while the diagram in (b) contributes to both the LP and NLP hard parts.   The cut diagram in (c) gives the LO gluon to heavy quark pair fragmentation function.}
\label{fig:hardpart}
\end{figure}

For the hard parts at NLP, we only need to calculate the rate to produce a heavy quark pair with zero relative transverse momentum, since the effect of relative transverse momentum will be further suppressed in $1/p_T$.   The hard parts can be perturbatively calculated order-by-order in powers of $\alpha_s$ by applying the factorization formula in Eq.~(\ref{eq:pqcd_fac}) to the production of an asymptotic state of a heavy quark pair $[Q\bar{Q}(\kappa)]$ of momentum $p$ with various spin and color quantum numbers.  For example, the hard part for the subprocess, $q+\bar{q}\to [Q\bar{Q}(\kappa)]+g$, can be derived by applying Eq.~(\ref{eq:pqcd_fac}) to heavy pair production in quark-antiquark scattering, 
\begin{eqnarray}
d\hat{\sigma}^{(3)}_{q+\bar{q}\to [Q\bar{Q}(\kappa)]+g}(p)
&=&
d\sigma^{(3)}_{q+\bar{q}\to [Q\bar{Q}(\kappa)]+g}(p)
\label{eq:qq2QQg}\\
&\ & \hskip -1in
- d\hat{\sigma}^{(2)}_{q+\bar{q}\to g+g}(p_g=p/z)
\otimes {\cal D}^{(1)}_{[Q\bar{Q}(\kappa)]/g}(z,m_Q)\, ,
\nonumber
\end{eqnarray}
where the superscript ``(n)'' indicates the order in $\alpha_s$.  The first term on the right of Eq.~(\ref{eq:qq2QQg}) is the differential cross section for the subprocess given by the diagrams in Fig.~\ref{fig:hardpart}(a) and (b) plus three more diagrams.  When $p_T\gg m_Q$, the square of diagram (a) clearly contributes to the NLP, while the the square of diagram in (b) contributes to both LP and NLP.  We note that the interference between the two contributes only to NLP.  The second term in Eq.~(\ref{eq:qq2QQg}) exactly removes the LP piece of this subprocess, where $d\hat{\sigma}^{(2)}_{q+\bar{q}\to g+g}(p_g)$ is the LO differential cross section for $q+\bar{q}\to g(p_g)+g$ and ${\cal D}^{(1)}_{[Q\bar{Q}(\kappa)]/g}(z,m_Q)$ is the LO fragmentation function given by the (cut) diagram in Fig.~\ref{fig:hardpart}(c).  Since the second term on the right of Eq.~(\ref{eq:qq2QQg}) removes a power mass singularity $\sim 1/m_Q^2$, it is important to keep the heavy quark mass when evaluating partonic diagrams and to set $m_Q \to 0$ (only) after carrying out the subtraction \cite{KQSfac:2011}.

The predictive power of the factorization formula in Eq.~(\ref{eq:pqcd_fac}) relies on the universality of the fragmentation functions. The single parton fragmentation functions $D_{H/f}(z,m_Q)$ are defined in Ref.~\cite{Nayak:2005rw}.  The operator definition for ${\cal D}_{H/[Q\bar{Q}(\kappa)]}(z,\zeta,\zeta',m_Q)$ depends on $\kappa$, which represents the pair's color and spin.  For the heavy quark pair moving in the ``$+z$'' direction with light-cone momentum components, $p^\mu=(p^+, (2m_Q)^2/2p^+,0_\perp)$, there are singlet ($1$) and octet ($8$) color states, and four spin states described by relativistic Dirac spin projection operators: $\gamma^+\gamma_5/4p^+$, $\gamma^+/4p^+$, and $\gamma^+\gamma^i/4p^+$ with $i=1,2$, for effective axial vector ($a$), vector ($v$), and tensor ($t$) ``currents'', respectively \cite{KQSfac:2011}.  As an example, the operator definition of the axial vector/octet fragmentation function can be written as 
\begin{eqnarray}
{\cal D}_{H/[Q\bar{Q}(a8)]}(z,\zeta,\zeta')
&=&
\sum_X \int \frac{p^+dy^- }{2\pi} \,
{\rm e}^{-i(p^+/z)y^-}
\nonumber\\
&& \hspace{-1.4in}\times 
\int \frac{p^+dy_1^- p^+dy_2^-}{(2\pi)^2}
{\rm e}^{i(p^+/2z)(1-\zeta)y_1^-}
{\rm e}^{-i(p^+/2z)(1-\zeta')y_2^-}
\nonumber\\
&&  \hspace{-1.4in} \times 
\frac{4}{(N^2-1)}
 \langle 0|\, \psi_i(0)\frac{\gamma^+\gamma_5}{4p^+}\, (t^a)_{ij}\, 
 \overline{\psi}_j(y_2^-) 
  | H(p^+)X\rangle
 \nonumber \\
&&  \hspace{-1.4in} \times  
 \langle H(P^+)X|
\psi_l(y^-+y_1^-)\frac{ \gamma^+\gamma_5}{4p^+}\, (t^a)_{lk}\,
\overline{\psi}_k(y^-)\, |0\rangle\, ,
\label{eq:frag_a8}
\end{eqnarray}
where we have suppressed dependence on a factorization scale.   We also suppress gauge links along the minus light cone, inserted between repeated color indices, which provide a gauge invariant definition of the operator \cite{KQSfac:2011}.     The links are in adjoint representation for index $a$, and fundamental representation for $i,j,l,k$.  (We note that the color matrices $t^a$ may be taken at any points along the light cone.)   Overall, the pair fragmentation functions are given by matrix elements of nonlocal operators, and the form is very similar to operator definitions of single parton fragmentation functions, simply replacing the parton field by the product of quark fields.   They are also reminiscent of hadronic wave functions that connect multiple partons to the hard scattering in the factorized expressions for elastic amplitudes \cite{Brodsky:1989pv}.

Similar to the single parton fragmentation functions, heavy quark pair fragmentation functions like the one defined in Eq.~(\ref{eq:frag_a8}) are nonperturbative, but, universal.  They are boost invariant, and require renormalization.   They thus evolve in the usual sense, and depend upon a factorization scale, chosen to match the short-distance scale of the problem.   It is natural to think of the choice of factorization scale as the same for NLP as for LP, and in general, a change of scale could mix them.   This leads to a closed set of general evolution equations for both LP and NLP fragmentation functions, which will be addressed elsewhere \cite{KQSfac:2011}.

We note that there are further additive corrections at the power of $1/p_T^2$, such as those involving twist-4 parton distributions, or twist-4 light-parton fragmentation functions.  These corrections, however, can be considered small since they do not introduce natural enhancements that scale with $m_Q$ in the probability to form a quarkonium,  to compensate the suppression of $1/p_T^2$.  

%

\noindent
{\it{ Cross section and polarization}}\,---\,It is the fragmentation functions that determine the absolute normalization of perturbative calculations using the factorization formula in Eq.~(\ref{eq:pqcd_fac}), as well as the differences in the production rate between various quarkonium states and their polarizations.  Since only pairs with small relative momentum are likely to form bound quarkonia, we may apply the basic NRQCD factorization hypothesis for heavy quarkonium production to these fragmentation functions to reduce the unknown functions to a few universal constants in the form of NRQCD matrix elements,
\begin{eqnarray}
&&D_{H/f}(z,m_Q,\mu)
= \sum_c 
d_{f\to[Q\bar{Q}(c)]}(z,m_Q,\mu)\langle O_{[Q\bar{Q}(c)]}^{H}\rangle
\nonumber\\
&&
{\cal D}_{H/[Q\bar{Q}(\kappa)]}(z,\zeta,\zeta',m_Q,\mu)\, ,
\nonumber\\
&& \
= \sum_c 
d_{[Q\bar{Q}(\kappa)]\to[Q\bar{Q}(c)]}(z,\zeta,\zeta',m_Q,\mu)\langle O_{[Q\bar{Q}(c)]}^{H}\rangle\, ,
\label{eq:frag_nrqcd}
\end{eqnarray}
where the functions $d$ are calculable, $\mu$ is a factorization scale, and $\langle O_{[Q\bar{Q}(c)]}^{H}\rangle$ are local NRQCD matrix elements \cite{Bodwin:1994jh}. Although we cannot provide a full proof for the NRQCD factorization in Eq.~(\ref{eq:frag_nrqcd}),  it is reasonable 
to evaluate the coefficient functions in Eq.~(\ref{eq:frag_nrqcd}) to estimate the properties of fragmentation functions \cite{Nayak:2005rw}.

\begin{figure}[!htp]
\noindent
\begin{minipage}[c]{0.4\columnwidth}
\noindent
        \includegraphics[width=1\columnwidth]{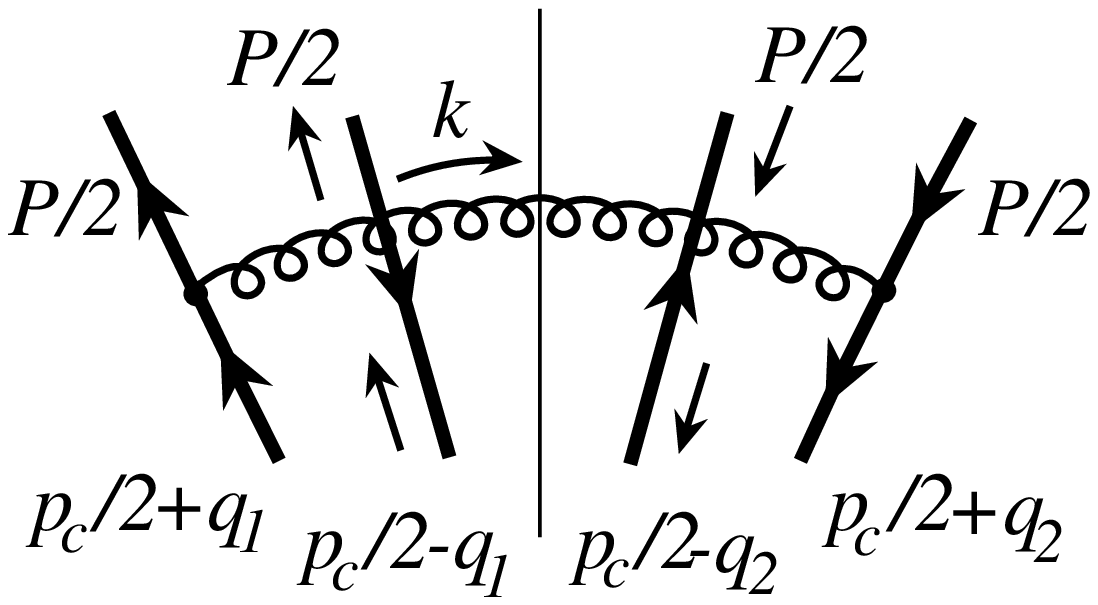}
\end{minipage}
\begin{minipage}[c]{0.03\columnwidth}
+
\end{minipage}
\begin{minipage}[c]{0.4\columnwidth}
\noindent
        \includegraphics[width=\columnwidth]{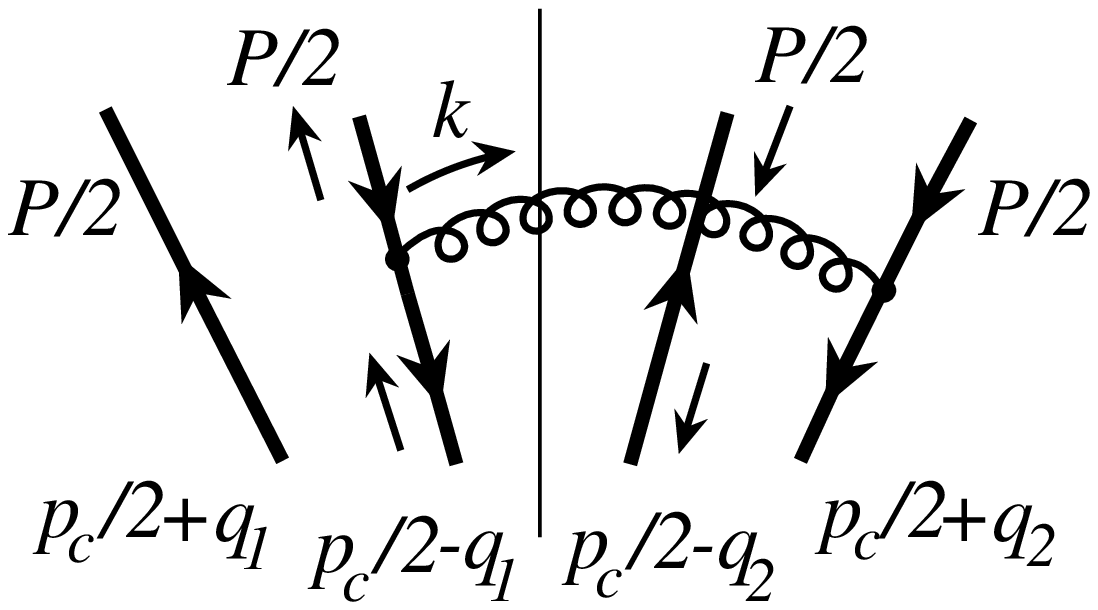}
\end{minipage}
\begin{minipage}[c]{0.08\columnwidth}
+ ...
\end{minipage}
\caption{Leading order Feynman diagrams represent the fragmentation of a heavy quark pair to another heavy quark pair.}
\label{fig:fragfunc}
\end{figure}

\begin{figure}[!htp]
\centering
\vskip 0.1in
        \includegraphics[width=0.8\columnwidth]{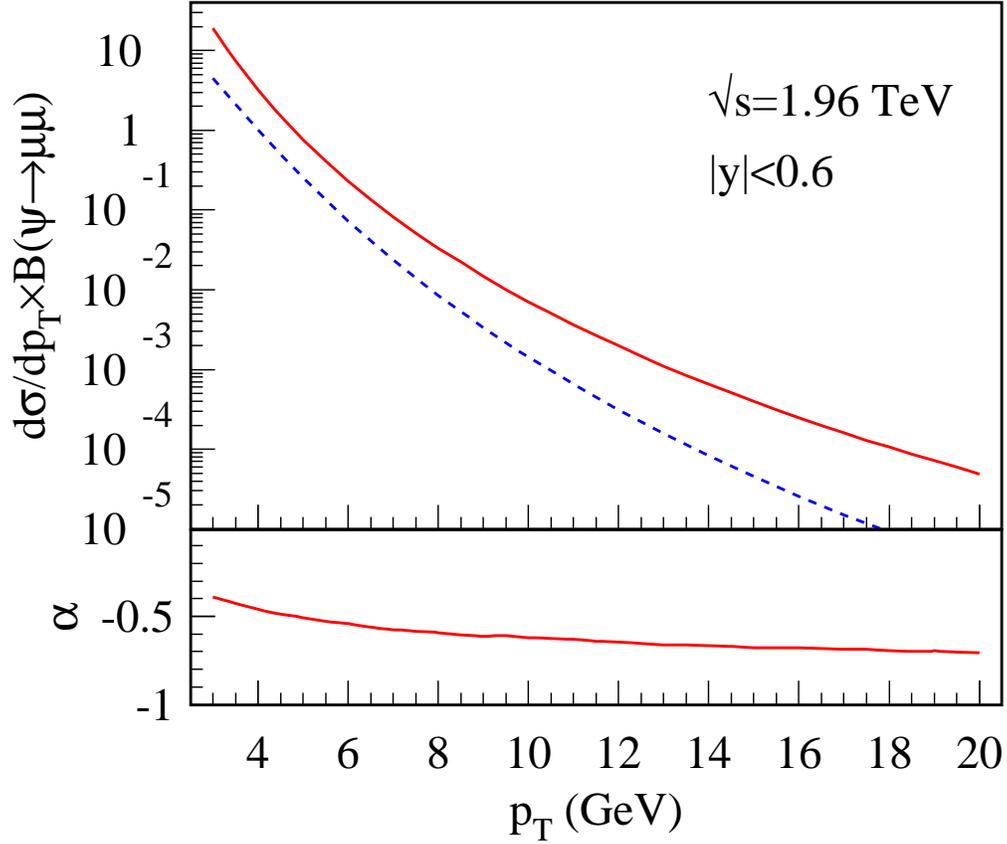}
\caption{$J/\psi$ cross section (upper panel) and polarization (lower panel) as a function of $p_T$.   The solid lines are computed from the NLP term in Eq.\ (\ref{eq:pqcd_fac}) with fragmentation functions given by Eq.\ (\ref{eq:fragpol}); the dashed line is LO CSM cross section as in Ref.\ \cite{Butenschoen:2010rq}, and all in [nb/GeV]. }
\label{fig:compare}
\end{figure}

We can use this formalism to help understand the source of the surprisingly large corrections to $J/\psi$ production in the CSM  at NLO and NNLO with predominantly longitudinal polarization, in contrast to the small and transversely polarized LO \cite{Gong:2008sn,Lansberg:2008gk}.
To do so, we use Eq.~(\ref{eq:pqcd_fac}) to calculate NLP $J/\psi$ cross sections at LO in $\alpha_s$ from a color singlet $^3{\rm S}_1$ pair.  We estimate the ${\cal O}(\alpha_s)$ fragmentation functions from the diagrams in Fig.\ \ref{fig:fragfunc}, where the upper lines are fixed at $p/2$ and Dirac indices are contracted with an NRQCD singlet projection with matrix element essentially equivalent to the CSM.  At order of $\alpha_s$, only the $[Q\bar{Q}(a8)]$ state fragments into a color singlet $^3{\rm S}_1$ heavy quark pair, while both $[Q\bar{Q}(v8)]$ and $[Q\bar{Q}(t8)]$ states give vanishing contributions due to charge conjugation symmetry \cite{KQSfac:2011}.  Defining $r(z) \equiv z^2\mu^2/(4m_c^2(1-z)^2)$, we have
\begin{eqnarray}
{\cal D}^L_{[Q\bar{Q}(a8)]\to J/\psi}(z,\zeta,\zeta',m_Q,\mu)
&=&
\frac{1}{2N^2}\frac{\langle O^{J/\psi}_{1(^3{\rm S}_1)}\rangle}{3m_c}
\Delta(\zeta,\zeta')
\nonumber\\
&& {\hskip -1.5 in}\times
\frac{\alpha_s}{2\pi}z(1-z)
\left[
\ln\left(r(z) + 1\right) 
- \left(1- \frac{1}{1+r(z)}\right)
\right],
\nonumber\\
{\cal D}^T_{[Q\bar{Q}(a8)]\to J/\psi}(z,\zeta,\zeta',m_Q,\mu)
&=&
\frac{1}{2N^2}\frac{\langle O^{J/\psi}_{1(^3{\rm S}_1)}\rangle}{3m_c}
\Delta(\zeta,\zeta')
\nonumber\\
&& {\hskip -1.5 in}\times
\frac{\alpha_s}{2\pi}z(1-z)\left[
1- \frac{1}{1+r(z)}
\right],
\label{eq:fragpol}
\end{eqnarray} 
where ${\cal D}^L$ (${\cal D}^T$) is the pair-fragmentation function with the pair longitudinally (transversely) polarized, and 
\begin{eqnarray}
\Delta(\zeta,\zeta')=\frac{1}{4}\sum_{a,b}
\delta(\zeta-a(1-z))\delta(\zeta'-b(1-z)),
\end{eqnarray}
with $a,b=-1,1$, and total unpolarized contribution, ${\cal D}_{[Q\bar{Q}(a8)]\to J/\psi}=2\,{\cal D}^T_{[Q\bar{Q}(a8)]\to J/\psi}+{\cal D}^L_{[Q\bar{Q}(a8)]\to J/\psi}$.  In deriving Eq.~(\ref{eq:fragpol}), we renormalized the UV divergence by a cutoff $\mu$ on transverse momenta.  Quark masses make an IR cutoff on transverse momenta unnecessary. Other renormalization schemes give similar results and will be discussed elsewhere \cite{KQSfac:2011}.  With these fragmentation functions, and LO hard scattering functions \cite{KQSfac:2011}, we can compute NLP cross sections that can be compared to NLO CSM cross sections  \cite{Campbell:2007ws,Lansberg:2008gk,Butenschoen:2010rq,Ma:2010yw}.  Using the CTEQ6L parton distributions \cite{Pumplin:2002vw}, $\mu=p_T$ for factorization and renormalization scales, $m_c=1.5$~GeV, and $\langle O^{J/\psi}_{1(^3{\rm S}_1)}\rangle=1.32$~GeV$^3$, we calculate and plot the unpolarized cross section as a function $p_T$ in Fig.~\ref{fig:compare} (upper panel).  We see that the factorization-based cross section shows much the same enhancement above the LO CSM as the full NLO result in CSM.   
Actually, our calculated NLP cross section with the estimated ${\cal O}(\alpha_s)$ fragmentation functions reproduces as much as eighty percent of the full NLO CSM result.  The difference should be due  mainly to ${\cal O}(\alpha_s)$ corrections to the hard scattering functions from the color singlet $[Q\bar{Q}(v1)]$ channel.  With  higher order fragmentation functions from other $[Q\bar{Q}(\kappa)]$ states, the full NLP cross section at LO in $\alpha_s$ could be larger than the solid line in Fig.~\ref{fig:compare}.  Similarly, we calculate the $J/\psi$ polarization (lower panel), 
as measured by the parameter  $\alpha=(\sigma^T-\sigma^L)/(\sigma^T+\sigma^L)$ in terms of  transverse (longitudinal) cross section $\sigma^T (\sigma^L)$.  Our result is consistent with that in Refs.\ \cite{Gong:2008sn,Lansberg:2008gk}.   We regard these results as compelling evidence for the phenomenological relevance of the power expansion.

%

\noindent
{\it{ Summary}}\,---\,We have argued that a practical strategy for the phenomenology of heavy quarkonium production at high $p_T$ is to expand the cross sections first in the large scale $p_T$, and then in the coupling, $\alpha_s$, and have presented a new perturbative QCD factorization formalism for  quarkonium production including the first non-leading powers in $m_Q/p_T$.  This approach enables us to resum perturbative logarithms into the fragmentation functions, to analyze systematically the influence of the larger color singlet matrix elements despite their suppressed $p_T$-dependence, and to resolve some of the mystery associated with the discovery of large high order corrections to color singlet cross sections.  We have found that heavy quarkonia produced from pair fragmentations are likely to be longitudinally polarized, in contrast to single parton fragmentation.  The observed quarkonium polarization should be a consequence of the competition of these two leading production channels.  

%

We thank G.T. Bodwin, E. Braaten, J.-P. Lansberg, B. Pire, and O.\ Teryaev for helpful discussions.
This work was supported in part by the U.S. Department of Energy, contract number DE-AC02-98CH10886, and by the National Science Foundation, grants PHY-0653342 and PHY-0969739.

\end{document}